\documentclass[lettersize,journal]{IEEEtran}
\usepackage{amsmath,amsfonts}
\usepackage{algorithmic}
\usepackage{array}
\usepackage[caption=false,font=normalsize,labelfont=sf,textfont=sf]{subfig}
\usepackage{textcomp}
\usepackage{stfloats}
\usepackage{url}
\usepackage{verbatim}
\usepackage{graphicx}
\def\BibTeX{{\rm B\kern-.05em{\sc i\kern-.025em b}\kern-.08em
    T\kern-.1667em\lower.7ex\hbox{E}\kern-.125emX}}
\usepackage{balance}
\begin{document}
\title{A Spatial-Temporal Progressive Fusion Network for Breast Lesion Segmentation in Ultrasound Videos}
\author{Zhengzheng Tu, Zigang Zhu, Yayang Duan, Bo Jiang, Qishun Wang and Chaoxue Zhang
\thanks{Corresponding authors: Bo Jiang and Chaoxue Zhang}
\thanks{Zhengzheng Tu, Zigang Zhu, Bo Jiang and Qishun Wang are with the Anhui Provincial Key Laboratory of Multimodal Cognitive Computing, School of Computer Science and Technology, Anhui University, Hefei, 230601, China (e-mail: zhengzhengahu@163.com, 18355421540@163.com, jiangbo@ahu.edu.cn, qishunahu@163.com).}
\thanks{Yayang Duan and Chaoxue Zhang are with the First Affiliated Hospital of Anhui Medical University, Hefei, 230022, China (drduan\_yayang@163.com, zcxay@163.com).}
}

\markboth{Journal of \LaTeX\ Class Files,~Vol.~18, No.~9, September~2020}%
{How to Use the IEEEtran \LaTeX \ Templates}

\maketitle

\begin{abstract}
Ultrasound video-based breast lesion segmentation provides a valuable assistance in early breast lesion detection and treatment. 
However, existing works mainly focus on lesion segmentation based on ultrasound breast images which usually can not be adapted well to obtain desirable results on ultrasound videos. 
%In contrast to images,  
The main challenge for ultrasound video-based breast lesion segmentation is how to exploit the lesion cues of both intra-frame and inter-frame simultaneously.
To address this problem, we propose a novel Spatial-Temporal Progressive Fusion Network (STPFNet) for video based breast lesion segmentation problem. 
The main aspects of the proposed STPFNet are threefold. 
First, we propose to adopt a unified network architecture to capture both spatial dependences within each ultrasound frame and temporal correlations between different frames together for ultrasound data representation. 
Second, we propose a new fusion module, termed Multi-Scale Feature Fusion (MSFF), to fuse spatial and temporal cues together for lesion detection. MSFF can help to determine the boundary contour of lesion region to overcome the issue of lesion boundary blurring. 
Third, we propose to exploit the segmentation result of previous frame as the prior knowledge to suppress the noisy background and learn more robust representation. 
In particular, we introduce a new publicly available ultrasound video breast lesion segmentation dataset, termed UVBLS200, which is specifically dedicated to breast lesion segmentation. It contains 200 videos, including 80 videos of benign lesions and 120 videos of malignant lesions. 
Experiments on the proposed dataset demonstrate that the proposed STPFNet achieves better breast lesion detection performance than state-of-the-art methods.
\end{abstract}


\section{Introduction}
\IEEEPARstart{B}{reast} cancer continues to be one of the most prevalent diseases for women.~\cite{b1}.
Many exploratory techniques have emerged in the field of breast cancer diagnosis and treatment~\cite{b2,b3}.
Among them, ultrasound has emerged as an important technique for breast lesion detection and diagnosis due to its superior ability to distinguish early indications~\cite{b4,b5}.
Segmentation of breast ultrasound images and detection of lesion areas are important steps to help physicians identify different functional tissues and guide tumour localisation and clinical treatment~\cite{b6}. 
\begin{figure}[htbp]
    \centering
    \includegraphics[width=8.5cm,height=5.5cm]{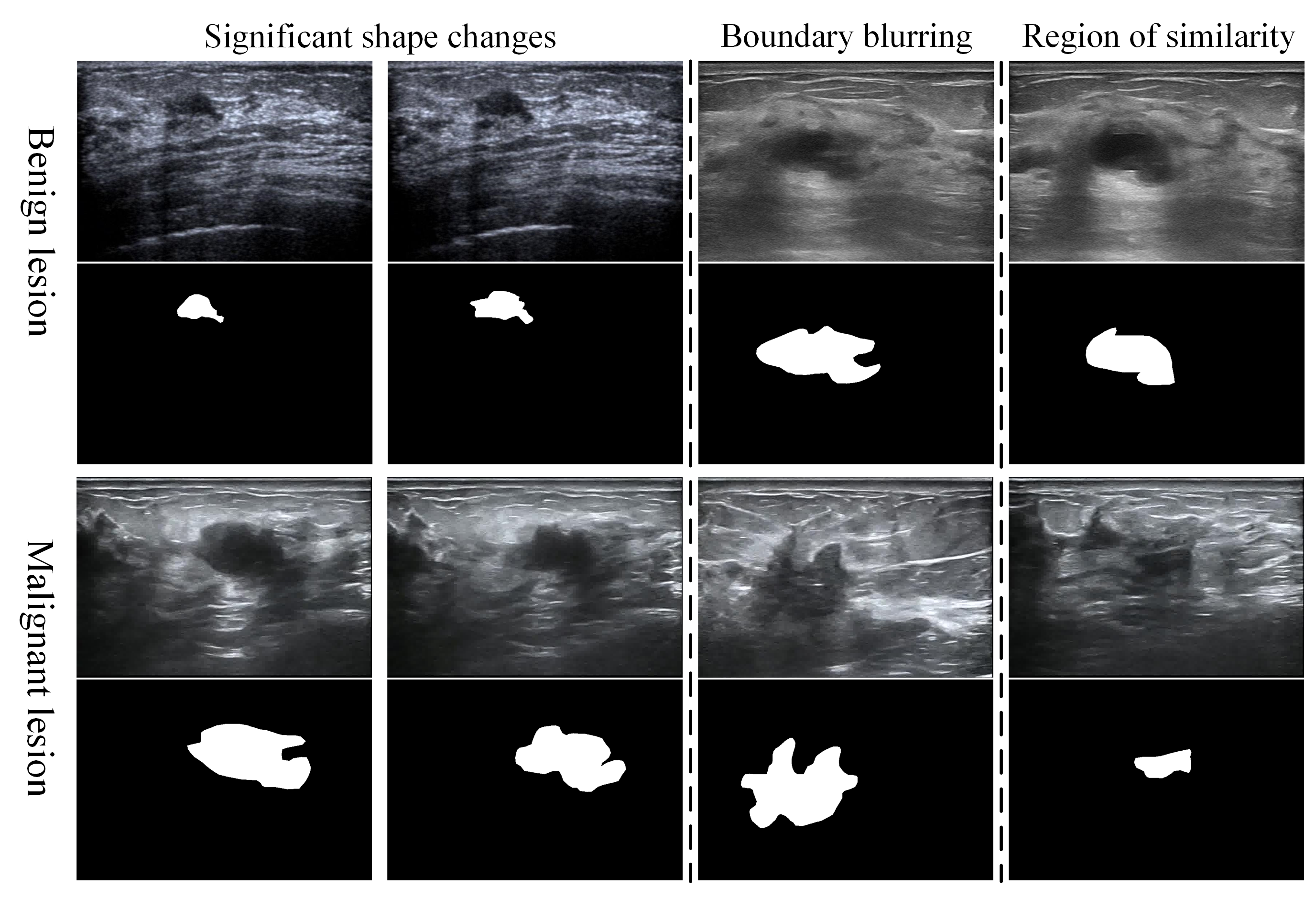}
    \caption{In the figure the first two rows represent benign lesions and their ground truth, and the last two rows represent malignant lesions and their ground truth. From left to right are the three types of challenges in our dataset, which are the significant change in shape of the lesions, the blurring of the boundaries, and the presence of regions similar to the lesions, respectively.}
    \label{fig1}
\end{figure}
However, different from natural images and other medical image segmentation, ultrasound data is usually low resolution, noisy, and similar in foreground and background~\cite{b7,b8,b9}.
These bring great challenges to the lesion segmentation tasks in ultrasound data, such as blurred tumor boundaries, and irregular shapes~\cite{b10}. 

To address the issues of boundary blurring of the myotonic junction (MTJ) in ultrasound images, Zhou \emph{et al.}~\cite{b11} proposed a Region-Adaptive Network (RAN) to locate MTJ region and segment it. It first employs a region-based multi-task learning module to explore regions containing MTJ and then extracts MTJ structures from the adaptively selected region via U-shaped paths to remove boundary ambiguities. 
Also, to address the issues of low resolution with thyroid nodules and shadow interference, Ouahabi \emph{et al.}~\cite{b12} proposed a fully convolutional dense dilation network which combines low-level fine segmentation with high-level coarse segmentation to extract more features from the ultrasound images.  
In addition to the above studies in the field of ultrasound, there have been some studies on addressing  ultrasound breast  segmentation problems.
Some researchers consider to conduct both tumour classification and segmentation together to achieve robust segmentation purpose. 
Zhou \emph{et al.}~\cite{b13} demonstrated that joint learning of these two tasks could help to improve their performance respectively. They propose a new multi-task learning scheme for accurate segmentation and classification of breast tumours. 
Also, some works improve the blurred boundary in segmentation by enhancing the learning of the blurred boundaries. 
For example, Wang \emph{et al.}~\cite{b14} designed a supervised residual representation module to learn ambiguous boundaries and confused regions by employing a residual feedback transmission strategy.
However, this residual network relies heavily on searching ambiguous boundaries when extracting features.
Some other strategies also consider to obtain more detailed features to reduce the problems associated with ultrasound imaging. 
For example, Xue \emph{et al.}~\cite{b15} generated a multilayer integrated feature map to learn remote non-local dependencies by integrating features from all CNN~\cite{b16} layers, and achieved accurate segmentation of breast lesion boundaries by learning additional boundary information.

In contrast to ultrasound images, ultrasound videos can provide richer spatial and temporal information for  lesion detection. 
Li \emph{et al.}~\cite{b17} presented a dataset and benchmark dedicated to segmentation of breast lesions in ultrasound videos.
In their work, the temporal transformer module focuses on motion information within the same region among consecutive frames, while the spatial transformer module makes full use of the information from the previous frame to capture the information spatially. Although the spatial-temporal information in the video is exploited in work~\cite{b17}, the above mentioned issues of ultrasound imaging, such as noise and blurred boundary, haven't been fully addressed. In addition, Lin \emph{et al.}~\cite{b44} proposed a larger dataset of ultrasound video breast lesions and a new Frequency and Localisation Feature Aggregation Network (FLA-Net) from a frequency domain perspective.

To overcome the above issues, in this paper, we propose a novel Spatial-Temporal Progressive Fusion Network (STPFNet) for video based breast lesion segmentation problem. The proposed STPFNet contrains three main aspects. First, STPFNet adopts a unified network architecture to capture both spatial dependences within each ultrasound frame and temporal correlations between different frames together for ultrasound data representation. Thus, it can fully exploit the cues of intra-frame and inter-frame simultaneously for accurate lesion segmentation. 
Second, in STPFNet, we design a new fusion module, termed Multi-Scale Feature Fusion (MSFF), to fuse spatial and temporal cues with different scales together for lesion detection. MSFF can help to determine the boundary contour of lesion region to overcome the issue of lesion boundary blurring in ultrasound videos. 
Third, STPFNet fully exploits the segmentation result of previous frame as the prior knowledge to suppress the noisy background and highlight the foreground to learn more accurate visual representation for lesion objects. 
In particular, we construct a novel dataset, termed UVBLS200, specifically designed for breast lesion segmentation tasks. The dataset consists of 80 video sequences depicting benign lesion and 120 video sequences depicting malignant lesion. Each video sequence contains the corresponding image frames along with the ground truth of the lesion. 
In Fig.~\ref{fig1}, we show our proposed dataset UVBLS200, which includes both benign and malignant lesions and their annotated ground truth. In addition, we briefly show some challenges in the UVBLS200 dataset, such as significant variation in lesion shape, blurred boundaries, and the presence of areas that are similar to the lesions.

Overall, the main contributions of this paper are summarized as follows, 

(1) We propose a new approach, called Spatial-Temporal Progressive Fusion Network (STPFNet), for breast lesion segmentation tasks. It makes full use of spatial-temporal information for ultrasound video via temporal and spatial fusion modules. 

(2) We design a new Multi-Scale Feature Fusion module, termed MSFF, to extract multi-scale (coarse-to-fine) information effectively for ultrasound video data representation. 

(3) We construct a publicly available ultrasound video breast lesion segmentation dataset (UVBLS200) which contains 200 video sequences totalling 10,666 images with ground truth annotations. The dataset will facilitate breast lesion segmentation studies and will be released to the public. % for free academic usage. 

(4) We evaluate our STPFNet method on UVBLS200 dataset on both image and video levels. The experimental results show that our proposed STPFNet achieves better performance compared to existing methods. 

\section{Related Work}
\subsection{Segmentation of breast lesions}
Ultrasound breast lesion segmentation plays a crucial role in the detection and treatment of breast lesions.
Formerly, ultrasound breast lesion segmentation relied primarily on the experience and observation of physicians. 
However, this way lacks reliability due to its subjectivity and inconsistency. 
Then, some attentions have been given to traditional image processing techniques and many methods~\cite{b18,b19,b20} have been developed in this field. 
These techniques to some extent facilitate the precise segmentation of ultrasound lesions.
With the booming development of machine learning technology, machine learning based breast lesion segmentation methods have emerged.
By exploiting the rich ultrasound image data and the autonomous learning capability of the algorithms, these methods~\cite{b21,b22,b23} learn a variety range of tumour features and achieve accurate tumour segmentation.
In contrast to traditional methods, these approaches have the ability to handle diverse shapes of lesion, thereby yielding more precise and stable segmentation results.
In recent years, along with deep learning being developed well, many deep neural network methods~\cite{b24,b25,b26,b27} have been proposed to segment breast lesions.
These methods usually involve multi-level feature extraction and incorporate contextual information to improve the accuracy of breast lesion segmentation.
Of these, U-Net~\cite{b28} plays a significant role in medical image segmentation.
It is composed of an encoder and a decoder interconnected through hopping connections, enabling better segmentation performance with less data. 
In addition, Soltani \emph{et al.}~\cite{b29} proposed a Mask R-CNN based method for accurate segmentation of multiple instances in an image by predicting both the bounding box and mask of the objects simultaneously.
Applying the benefits of deep learning to breast lesion segmentation is expected to provide physicians with more reliable assistance and achieve more accurate lesion segmentation.
\subsection{Video Object Segmentation}
Initially, the background and development of video object segmentation can be traced back to the early still image segmentation techniques. Subsequently, with the rise of video processing techniques, researchers began to propose various algorithms and methods for the dynamic characteristics of video. 
Wei \emph{et al.}~\cite{b30} proposed an automatic video segmentation algorithm based on k-median clustering algorithm and 2D binary model.
However, these methods almost lie on low-level feature representation, and rarely utilize high-level feature representation.
As a result, the object segmentation in complex scenes is relatively unsatisfactory. Semi-supervised learning has been incorporated into video object segmentation recently. 
In semi-supervised video object segmentation, instead of labelling the whole video in detail, only a part of the video sequence needs to be labelled. The model then learns the mapping relationship from labelled frames to unlabelled frames. Semi-supervised video object segmentation not only uses labelled training data, but also makes full use of unlabelled data to improve the performance of the model.
STM~\cite{b31} employed space-time fusion to integrate the information of past and current frames, and decoded it using skip connections, resulting in a considerable improvement in performance.
KMN~\cite{b32} presented an enhanced STM approach, for addressing the issue of object missegmentation by introducing memory-to-query matching.
Nevertheless, this solution relies only on pixel similarity matching and is limited in its ability to handle appearance changes and deformations.
Furthermore, RMNet~\cite{b33} mitigated the limitations of STM to some extent, by incorporating the mask predicted by the current frame with optical flow calculation.
\begin{figure*}[ht]
    \centering
    \includegraphics[width=18cm,height=8cm]{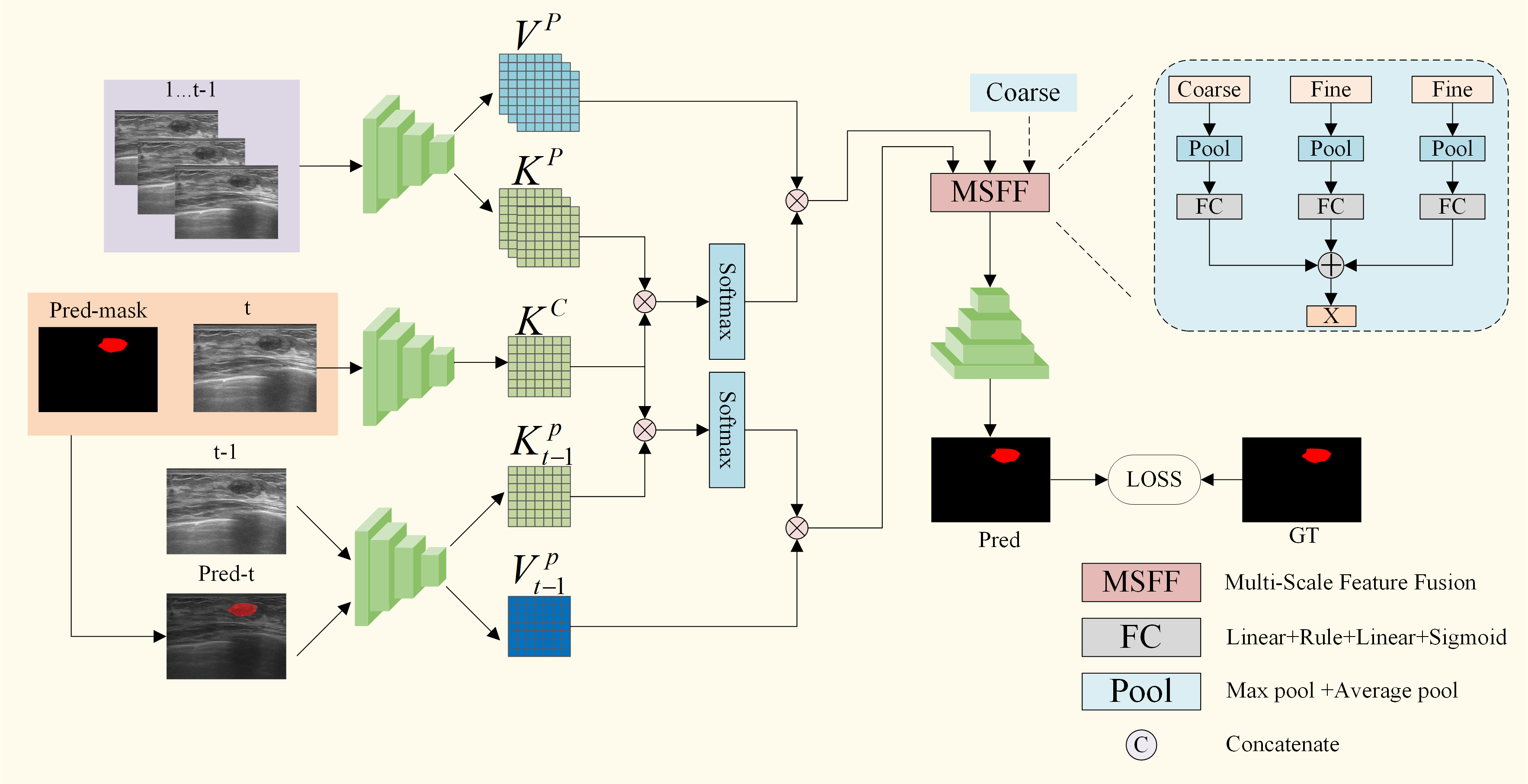}
    \caption{In our proposed network, we denote frames 1 to t-1 in the video as past frames. The term "pred-t" denotes the multiplication of the prediction mask from the previous frame with the current frame. "t-1" represents the previous frame. Subsequently, the "Pred-t" and the previous frame "t-1" are fed into the network backbone, thus generating $\mathbf{V}^{p}_{t-1}$ and $\mathbf{K}^{p}_{t-1}$ respectively. These obtained features are then fused together by the MSFF module, and the final prediction is produced by the decoder.}
    \label{fig2}
\end{figure*}

\section{Method}
This section provides a comprehensive overview of the Spatial-Temporal Progressive Fusion Network, including the Temporal Fusion Module (TFM), the Spatial Fusion Module (SFM), the Multi-Scale Feature Fusion Module (MSFF), and the loss function.
\subsection{Overall Architecture}
Fig.~\ref{fig2} illustrates the overall framework of the proposed Spatial-Temporal Progressive Fusion Network in this paper.
The Temporal Fusion Module, similar to STM~\cite{b31}, fuses the current frame with the past frames to find similar regions between the current and the past frames in the video sequence.
The Spatial Fusion Module fuses the current frame with the previous frame. It takes the predicted lesion region from the previous frame as a priori knowledge to suppress the noisy background and highlight the foreground to achieve the localisation of the lesion region.
Then to improve the performance, we further use a multi-scale feature fusion approach.
This involves the fusion of the ultrasound breast lesion information obtained by the encoder with the fusion information of the two branches described above.
Subsequently, the features obtained from the fusion are up-sampled by a decoder, resulting in a prediction that matches the size of the input image.
\subsection{Temporal Fusion Module}
In this work, we draw on video object segmentation methods developed for natural images to improve our approach.
Specifically, the space-time memory network skillfully uses the temporal information within a video sequence.

Inspired by STM~\cite{b31}, we use ResNet50~\cite{b34} as the encoder, and define the feature of the last layer of the current frame through the encoder as $\mathbf{r}\in\mathbb{R}^{H\times W\times C}$, where $H$ stands for the height of the feature, $W$ for the width of the feature, and $C$ represents the number of channels of the feature.
Without reducing the resolution, the number of channels of $\mathbf{r}$ is reduced to 1/2 and 1/8 of the original number of channels as the value and key of the current frame, defined as $\mathbf{V}^{C} \in \mathbb{R}^{H \times W \times C / 2}$ and $\mathbf{K}^{C}\in\mathbb{R}^{H\times W\times C/8}$, respectively.
For past frames, the same operation as above is performed to get the value and key for each past frame. Then all past frames are concatenated to form the value and key of all past frames, defined as $\mathbf{V}^{P} \in \mathbb{R}^{T \times H \times W \times C / 2}$, $\mathbf{K}^{P} \in \mathbb{R}^{T \times H \times W \times C / 8}$.
Then the similarity weight between the $\mathbf{K}^{C}$ of the current frame and the $\mathbf{K}^{P}$ of the past frames is calculated. Finally the obtained similarity weights are mapped on the past frames to find the most similar position between the current frame and the past frames. The similarity calculation function $f$ is shown below:
\begin{equation}
\begin{aligned}
\ f({K}^{C},{K}^{P})=softmax(exp({K}^{C}\otimes {K}^{P}))
\end{aligned}
\end{equation}
where $\otimes$ represents matrix multiplication.
By mapping the similarity weights to past frames, the final feature $y$ is obtained as follows:
\begin{equation}
\begin{aligned}
\ y=f({K}^{C},{K}^{P}){V}^{P}.
\end{aligned}
\end{equation}
Like the STM, the temporal fusion module has similar limitations.
Due to the noise and blurring in ultrasound video frames, the temporal fusion module still incorrectly segments similar regions as the lesion.
\subsection{Spatial Fusion Module}
Inspired by this work~\cite{b35}, we find that the object region in adjacent frames of video exhibits a certain level of continuity. 
This implies that the approximate position of the object region can be inferred from the frames with a certain degree of overlap. 
Therefore, by utilizing the positional information from the previous frame, more details of the lesion can be obtained.
The SFM can encode the previous neighbouring frame and add the lesion prediction results from the previous frame as a prior knowledge to locate the lesion more accurately.
\begin{figure}[ht]
    \centering
    \includegraphics[width=8.5cm]{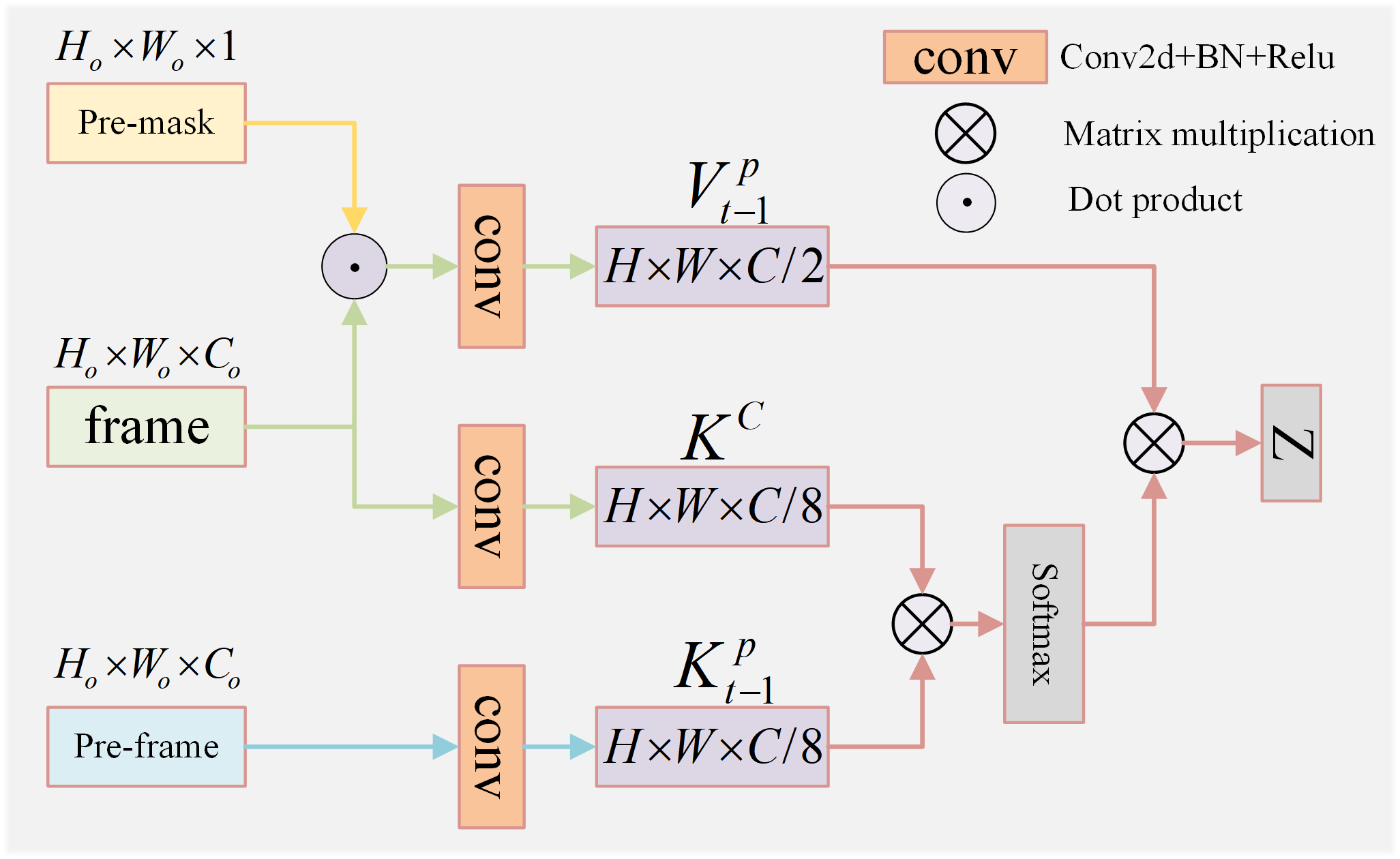}
    \caption{This is our spatial fusion module. The prediction of the previous frame ($H_{o}\times W_{o}\times 1$) is multiplied with the current frame ($H_{o}\times W_{o}\times C_{o}$) as a priori knowledge to suppress the background noise.}
    \label{fig3}
\end{figure}
As shown in Fig.~\ref{fig3}, we convolve the previous frame to obtain the corresponding key, defined as $\mathbf{K}^{p}_{t-1} \in \mathbb{R}^{H \times W \times C / 8}$, where $t-1$ denotes the previous frame of the current frame in the same video sequence. The previous frame has not only most similar texture information as the current frame in the spatial dimension, but also most close lesion location as the current frame in the temporal dimension. 
Then we perform a similarity calculation $f$ between the ${K}^{p}_{t-1}$ of the previous frame and the ${K}^{C}$ of the current frame. The calculation process is as follows:
\begin{equation}
\begin{aligned}
\ f({K}^{C},{K}^{p}_{t-1})=softmax(exp({K}^{C}\otimes {K}^{p}_{t-1}))
\end{aligned}
\end{equation}
where $\otimes$ represents matrix multiplication.
Furthermore, considering the overlap of lesion regions between frames, we multiply the lesion region predicted in the previous frame as a prior knowledge with the current frame, which can make the network to pay more attention to the lesion region predicted in the past and suppress the background noise.
The formula is as follows:
\begin{equation}
\begin{aligned}
\ {V}^{p}_{t-1}=\theta ({P}_{m}\ast {C}_{f})
\end{aligned}
\end{equation}
where ${P}_{m}$ indicates the predicted mask of the previous frame, ${C}_{f}$ indicates the current frame, $\theta$ means subsequent operations that are encoding and convolution, and ${V}^{p}_{t-1}$ denotes the value to which a priori knowledge is added.
By mapping the similarity weights onto the current frame, the final feature $z$ is obtained as follows:
\begin{equation}
\begin{aligned}
\ z=f({K}^{C},{K}^{p}_{t-1}){V}^{p}_{t-1}.
\end{aligned}
\end{equation}
\subsection{Multi-Scale Feature Fusion Module}
\begin{figure}[ht]
    \centering
    \includegraphics[width=8.5cm]{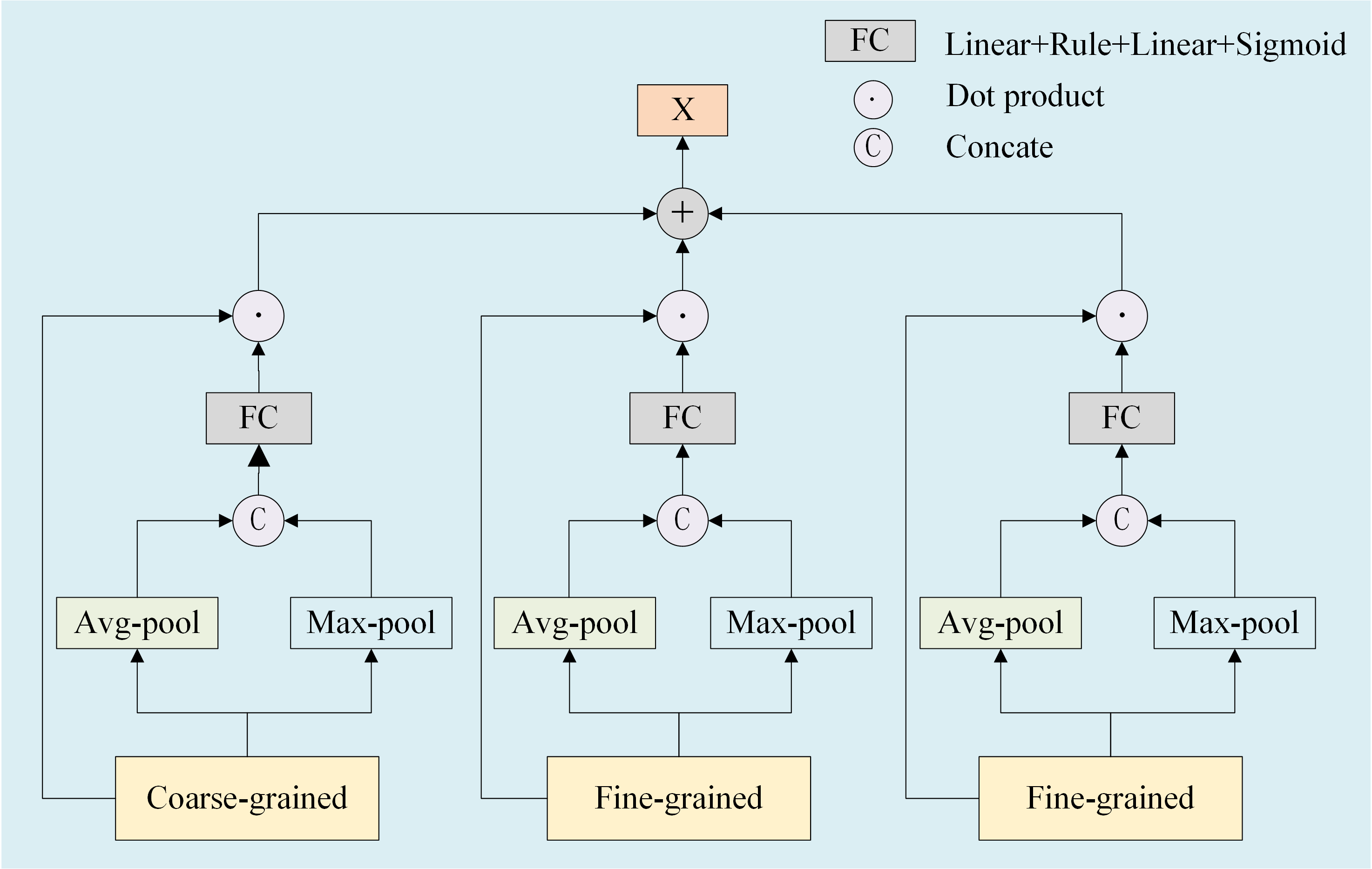}
    \caption{The multi-scale feature fusion module is proposed. Features obtained by spatial-temporal fusion are categorised as fine-grained information, while features obtained with convolution only are categorised as coarse-grained information. The features are fused using a weighted summation performed by two Linear layers.}
    \label{fig4}
\end{figure}

In ultrasound lesion segmentation, it is important to achieve fine segmentation of lesion boundaries. However, it is difficult to solve the problem of noise and blurred boundaries in ultrasound videos simply using the approach of video object segmentation. 
To address these issues, we propose a Multi-Scale Feature Fusion method to interactively fuse temporal, spatial, and coarse-grained features to realize the segmentation of boundaries. As shown in Fig.~\ref{fig4}.
Specifically, we take the features from two different scales as the fine-grained features, including the global similarity region features from the temporal fusion module and the lesion region features from the spatial fusion module. Furthermore, we take the features from the encoder as coarse-grained information and introduce other fine-grained information at multiple scales, making the network to learn more detailed features and achieve better contour information of the lesion. 
First, we denote the features obtained by the temporal fusion module, the spatial fusion module and the encoder as $y \in \mathbb{R}^{H \times W \times C / 2}$, $z \in \mathbb{R}^{H \times W \times C / 2}$ and $w \in \mathbb{R}^{H \times W \times C / 8}$, respectively. Since, average pooling is to obtain the average value in the local window, which has a certain smoothing effect and helps to reduce the influence of noise. Secondly, max pooling has certain invariance for small translation of object in input images. Even if the lesion moves slightly in the window, max pooling can still extract the same characteristic. In the multi-scale feature fusion module, we perform max pooling and average pooling on $z$, $y$, and $w$, respectively. And the results of the pooling will be concated and fed into the $FC$ to obtain the weight maps denoted as $Z$, $Y$, $W$, respectively. The above process can be described as follows:
\begin{equation}
\begin{aligned}
\ Y=\Phi (avg(y)\oplus max(y))
\end{aligned}
\end{equation}
\begin{equation}
\begin{aligned}
\ Z=\Phi (avg(z)\oplus max(z))
\end{aligned}
\end{equation}
\begin{equation}
\begin{aligned}
\ W=\Phi (avg(w)\oplus max(w))
\end{aligned}
\end{equation}
where $\oplus$ stands for concat operation, $\Phi$ represents the $FC$ which mainly consists of two Linear layers, $avg$ stands for average pooling, and $max$ stands for max pooling. 
we mutiply the weight maps $Z$, $Y$, and $W$ with $z$, $y$, and $w$ to get the weighted features and directly add them to obtain the mulit-scale fusion feature denoted as $X$. And the poccess can be described as:
\begin{equation}
\begin{aligned}
\ X = Y\cdot y+  Z\cdot z+  W\cdot w.
\end{aligned}
\end{equation}
Finally, we fed the mulit-scale fusion feature into the decoder to produce the prediction maps.
\subsection{Loss Function Calculation}
In this paper, we use three consecutive frames with their ground truth during the training process, and provide the ground truth of the first frame to predict the result of segmentation for the subsequent frames during the testing process. In the training phase, we use the cross-entropy loss function to calculate the distance between the prediction of our network and the ground truth. Here we denote the ground truth as $g2$,$g3$ and the predictions as $p2$, $p3$, respectively. The $CELoss$ is described as:
\begin{equation}
\begin{aligned}
\ \text CELoss(g2, p2) = - \sum g2 * log(p2) 
\end{aligned}
\end{equation}
\begin{equation}
\begin{aligned}
\ \text CELoss(g3, p3) = - \sum g3 * log(p3). 
\end{aligned}
\end{equation}
The total loss is formulated as:
\begin{equation}
\begin{aligned}
\ Loss = CELoss(g2, p2) +  CELoss(g3, p3).
\end{aligned}
\end{equation}

\begin{table*}[htbp]
\centering
   % \begingroup 
    \caption{Comparison with the state-of-the-art methods. The dataset is evaluated using both image-level and video-level methods, and the optimal results are highlighted in \textbf{bold}}
     \setlength{\tabcolsep}{19pt}
     \renewcommand{\arraystretch}{1.6} 
    \begin{tabular}{cccccccc}
    \hline
    \textbf{Method} & \textbf{Venue} & \textbf{Year}  & \textbf{Type} & \textbf{Dice} &  \textbf{Iou}  & \textbf{Recall} & \textbf{MAE} \\
    \hline
    Unet++   & TMI     & 2019  & Image  & 0.723 & 0.576  & 0.664 & 0.054 \\
    HarDNet  & ICCV    & 2019  & Image  & 0.823 & 0.727  & 0.823 &\textbf{0.035}\\
    MSNet    & MICCAI  & 2021  & Image  & 0.800 & 0.700  & 0.769 & 0.041\\
    TRUNet   & arxiv   & 2022  & Image  & 0.819 & 0.724  & 0.828 & 0.038 \\
    UCTNet   & AAAI    & 2022  & Image  & 0.825 & 0.721  & 0.846 & 0.037 \\
     \hline
    STM      & ICCV    & 2019  & Video  & 0.821 & 0.729  & 0.857 & 0.039 \\
    AFB-URR  & NeurIPS & 2020  & Video  & 0.811 & 0.713  & 0.794 & 0.037 \\
    DCF-Net  & ICCV    & 2021  & Video  & 0.804 & 0.707  & 0.794 & \textbf{0.035} \\
    UFO      & TMM   & 2022  & Video  & 0.789 & 0.680  & 0.813 & 0.040  \\
    \hline
    TMFF (Ours) &  -   & - & Video  &\textbf{0.841} &\textbf{0.752} &\textbf{0.888} &\textbf{0.035} \\
    \hline
    \end{tabular}%
     % \endgroup
  \label{tab1}%
\end{table*}%
 
\section{UVBLS200: Ultrasound Video Breast Lesion Segmentation Dataset}
%\label{sec:guidelines}
For the development of ultrasound video lesion segmentation, we construct a new ultrasound video breast lesion segmentation dataset, and we will introduce the dataset in detail in this section. 
\subsection{Data Acquisition}
Our ultrasound video breast lesion segmentation dataset (UVBLS200) was collected from the total of 528 patients, using Resona 7 and Toshiba 660a ultrasound systems and including 200 video sequences with a total of 10,666 frames. The video sequences have varying resolutions, with a maximum of 1072 × 756 and a minimum of 256 × 256. The data used in the dataset presented in this paper have been approved by the Clinical Medical Research Ethics Committee of the First Affiliated Hospital of Anhui Medical University. The UVBLS200 dataset does not contain any personal patient information.
\subsection{Dataset Description}
We describe UVBLS200 dataset in detail, which is formatted in DAVIS 2017 and consists of 200 video sequences with a total of 10,666 frames, where there are 80 benign and 120 malignant lesions in 200 video sequences. 
For the video sequences, the ground truth of each frame was annotated by three radiologists (at least 3 years of medical experience) and a senior department director was responsible for supervising and refining the quality of the ground truth to ensure the accuracy of our dataset. Two aspects of the main contribution of our dataset are as follows:

(1) We provide a publicly available dataset of ultrasound video breast lesion segmentation, annotated by specialized physicians to ensure the accuracy of our dataset.

(2) Our dataset is more realistic and comprehensive than the available published ultrasound breast datasets. Our data contains not only typical lesion samples, but also atypical samples that appear to be caused by a variety of factors, making the dataset more diverse. 

For example, some images have more blurred lesion boundaries and more pronounced changes in tumour shape than some typical images. In addition, since our dataset consists of video sequences, it is possible to observe the same lesion in multiple frames, thus obtaining more detailed information about the lesion.

\section{Experiment}
In this section, we will introduce the set up of the experiment and the evaluation metrics. In addition, we compare our method with the state-of-the-art methods. Finally, we conduct ablation experiments to verify the effectiveness of each proposed module.
\subsection{Experimental Detail}
We use 90\% of the video sequences in the UVBLS200 dataset for training and 10\% for testing.
Our network is implemented on an NVIDIA A100 GPU, utilizing Python 3.6 and PyTorch development environment 1.6.0.
\subsection{Evaluation Metrics}
To ensure a systematic and comprehensive evaluation of the experimental results, we employ four indicators to assess the segmentation results.
We utilize the $Dice$ coefficient which quantifies the proportion of overlapping region to measure the similarity between the segmentation result and the ground truth. To assess the degree of overlap between the segmentation results and the ground truth, we use the $IoU$ to calculate the ratio of the intersection to the union of both. In addition, we exploit $Recall$ to evaluate the capability of model for detecting true instance. Finally, we use $MAE$ to measure the error between the predicted value and the actual value. The smaller the value, the better the predictive power of the model.
The following formulas represent the specific implementation of four metrics:
\begin{equation}
\begin{aligned}
\ Dice = (2  |SR \cap GT|) / (|SR| + |GT|)
\end{aligned}
\end{equation}
\begin{equation}
\begin{aligned}
\ IoU = (|SR \cap GT|) / (|SR \cup GT|)
\end{aligned}
\end{equation}
where $|SR|$ represents the set of pixels in the segmentation result, and $|GT|$ represents the set of pixels in the ground truth.
$|SR \cap GT|$ denotes the number of pixels in the intersection of $SR$ and $GT$, $|SR \cup GT|$ denotes the number of pixels in the concatenation of $SR$ and $GT$.
\begin{equation}
\begin{aligned}
\ Recall = TP / (TP + FN)
\end{aligned}
\end{equation}
where $TP$ denotes the count of true positive examples, while $FN$ represents the count of false negative examples.
\begin{equation}
\begin{aligned}
\ MAE=\frac{1}{n} \sum_{i=1}^{n}\left|\hat{y}_{i}-y_{i}\right|
\end{aligned}
\end{equation}
where, $n$ represents the number of samples, $\hat{y}_{i}$ represents the actual value of the $i$ sample, and $ y_{i}$ represents the predicted value of the $i$ sample.
\subsection{Comparison with State-of-the-Art Methods}
\subsubsection{Comparison of Different Methods}  
In this section, we compare our method with state-of-the-art methods separately. The methods of images include: UNet++~\cite{b36}, HarDNet~\cite{b37}, MSNet~\cite{b38}, TRUNet~\cite{b39}, and UCTNet~\cite{b40}. The methods of videos include: STM~\cite{b31}, AFB~\cite{b41}, DCF-Net~\cite{b42}, and UFO~\cite{b43}. STM locates objects by finding the similarity region between past and current frames in a video. And AFB-URR further exploits the uncertainty region to refine the boundary based on STM. DCF-Net exploits spatial-temporal information by extracting the position-related affinities between consecutive frames. In addition, UFO changes the last two layers of the skip connection to a transformer block, making it easier to capture long-term dependencies between features. To ensure the fairness of experiments, we train all methods on the UVBLS200 dataset. The experimental results in Table~\ref{tab1}, we can see that our method outperforms other methods in all three metrics of dice, iou and recall, and achieves the best performance. Specifically, compared to the baseline~\cite{b31}, our method improves 2.0\%, 2.3\%, and 3.1\% on the Dice, Iou, and Recall metrics, respectively, while decreasing 0.4\% on the MAE. In addition, compared to the UCTNet network, we improve 1.6\%, 3.1\%, and 4.2\%, respectively, while decreasing 0.2\% in MAE. The methods for images also achieves good results on our dataset, indirectly proving the validity of our dataset. The analysis for the results shows that our method achieves satisfactory segmentation results for ultrasonic videos.
%~\ref{tab:tab1}
%Table generated by Excel2LaTeX from sheet 'Sheet1'
\begin{table}[htbp]
\centering
      \caption{The impact of two modules in the network on the dataset, on Dice, IoU, Recall and MAE metrics}
      \setlength{\tabcolsep}{7pt}
       \renewcommand{\arraystretch}{2} 
        \begin{tabular}{ccc|cccc}
        \hline
        \textbf{Baseline} & \textbf{MSFF} & \textbf{SFM} & \textbf{Dice} & \textbf{Iou}  & \textbf{Recall} & \textbf{MAE}  \\
         \hline
         $\surd $     &             &              & 0.821 & 0.729& 0.857& 0.039  \\
         $\surd $     & $\surd $    &              & 0.826 & 0.732& 0.842& 0.036 \\
         $\surd $     &             & $\surd $     & 0.832 & 0.744 & 0.888& 0.040   \\
         \hline
         $\surd $     & $\surd $    & $\surd $     & 0.841 & 0.752& 0.888& 0.035 \\
         \hline
         \end{tabular}%
       \label{tab2}%
\end{table}%

\begin{figure*}[ht]
\centering
    \includegraphics[width=17.5cm,height=12cm]{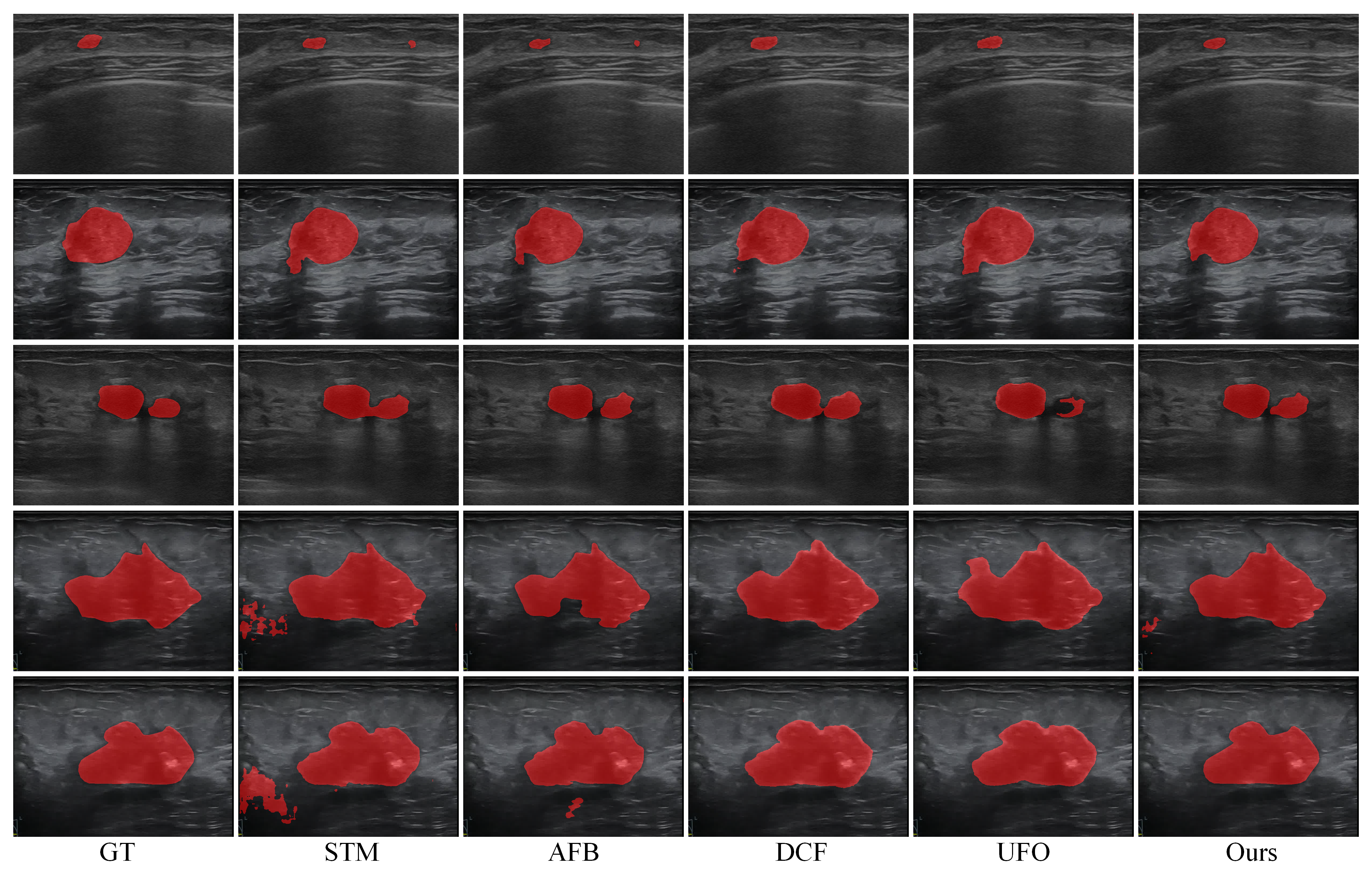}
    \caption{Visual presentation of comparative experiments. The first column corresponds to the ground truth, the second column shows the baseline prediction results, the third column shows the AFB prediction results, the fourth row shows the DCF prediction results, the fifth row shows the UFO prediction results, and the sixth column represents the prediction results of our method. Where the breast lesion area is highlighted in red.}
    \label{fig5}
\end{figure*}

\subsubsection{Analysis for Visual Results}
We focus on comparing the visualisation results of the methods for videos, as shown in Fig.~\ref{fig5}.
We can see our method performs a significant improvement for segmentation of lesion boundaries. 
In addition, due to the high noise in the ultrasound image, it is hard to locate the lesion region correctly, and to segment the region similar to the lesion (e.g., STM and AFB in the first and fifth rows) accurately. We use the previous frame as a prior knowledge, thus reducing the incorrect segmentation of the region similar to the lesion to a certain extent. For boundary blurring of ultrasound breast lesion, we use multi-scale feature fusion to extract more detailed information from different scales, which can locate the boundary contour more accurately than other methods.

\subsection{Ablation Study}
\subsubsection{Architecture Ablation}
We conduct a series of ablation experiments to demonstrate the effectiveness of the designed module.
In the spatial fusion module, considering that the previous frame and the current frame in a video usually have close location for the lesion, we apply the segmentation result of the previous frame as a prior knowledge to locate the lesion region of the current frame accurately.
In addition, we calculate the similarity between the current frame and the previous frame with the aim of minimising the cases of incorrect segmentation of the lesion region.
As shown in Table~\ref{tab2}, this modification resulted in improvements compared to the baseline, including a 0.9\% increase on Dice, a 0.8\% increase on IoU, and a 0.5\% decrease on MAE.

In the multiscale feature fusion module, we fuse the features obtained from the time fusion branch, the spatial fusion branch and the encoder to obtain richer detail information, which not only reduces information loss but also better complements the lesion boundaries.
During the ablation experiments, we simply merge the features obtained from the two fusion branches and compare the performance with the MSFF module to demonstrate the effectiveness of the MSFF module.
As shown in Table~\ref{tab2}, the multi-scale feature fusion module leads to a 1.5\% increase in the Dice coefficient, a 2.0\% increase in the IoU, a 4.6\% increase in Recall, and a 0.1\% decrease in MAE when compared to the baseline.
\begin{table*}[htbp]
\centering
     \caption{Ablation experiments. "+2", "+3", and "+4" represent the features encoded in the third last, second last, and last encoder layers, respectively. And "$\ast$mask" indicates the mapping method that maps the prediction result of the previous frame to the current frame}
      \setlength{\tabcolsep}{10.5pt}
       \renewcommand{\arraystretch}{1.7} 
        \begin{tabular}{c|cc|ccc|c|cccc}
        \hline
        \textbf{Baseline} & \textbf{+Maxpool} & \textbf{+Avgpool}  & \textbf{+2}& \textbf{+3}& \textbf{+4}&\textbf{$\ast$mask}& \textbf{Dice} & \textbf{Iou} & \textbf{Recall} & \textbf{MAE}\\
         \hline
        % $\surd $ &          &           &         &          &          &  $\surd $ &0.834 & 0.740 &  0.843\\
         $\surd $ & $\surd $ &           &         &          & $\surd $ &  $\surd $ &0.837 & 0.747 & 0.861 & 0.037\\
         $\surd $ &          & $\surd $  &         &          & $\surd $ &  $\surd $ &0.834 & 0.743 &  0.889& 0.039\\
         $\surd $ & $\surd $ & $\surd $  & $\surd $&          &          &  $\surd $ &0.836 & 0.745 & 0.881& 0.038\\
         $\surd $ & $\surd $ & $\surd $  &         &  $\surd $&          &  $\surd $ &0.834 & 0.742 &  0.852& 0.040\\
         $\surd $ & $\surd $ & $\surd $  &         &          & $\surd $ &           &0.821 & 0.726 &  0.810& 0.037\\
         $\surd $ & $\surd $ & $\surd $  &         &          & $\surd $ &  $\surd $ &0.841 & 0.752 & 0.888& 0.035\\
         \hline
         \end{tabular}%
       \label{tab3}%
\end{table*}%
\subsubsection{Parameter Ablation}

We also compare some key parameters and analyze the differences between them. As shown in Table~\ref{tab3}.
Specifically, we investigate the effectiveness of the strategy of mapping predictions from previous frames to the current frame in the spatial fusion module, and explore the effectiveness of adding coded information as coarse-grained information as well as the optimal coding layer. Finally, we compare the advantages of max and average pooling layers in multiscale feature fusion.
Based on the experimental results, simple multiplication of the previous frame prediction results as a priori knowledge with the current frame helps to suppress background noise and locate the lesion area. In addition, the features extracted from the different layers in the encoder contribute to the lesion segmentation, especially from the last layer.
The simultaneous use of combined max pooling and average pooling helps to smooth out feature variations and makes the network focus on the most critical regions.

\begin{figure}[ht]
\centering
    \includegraphics[width=8.5cm]{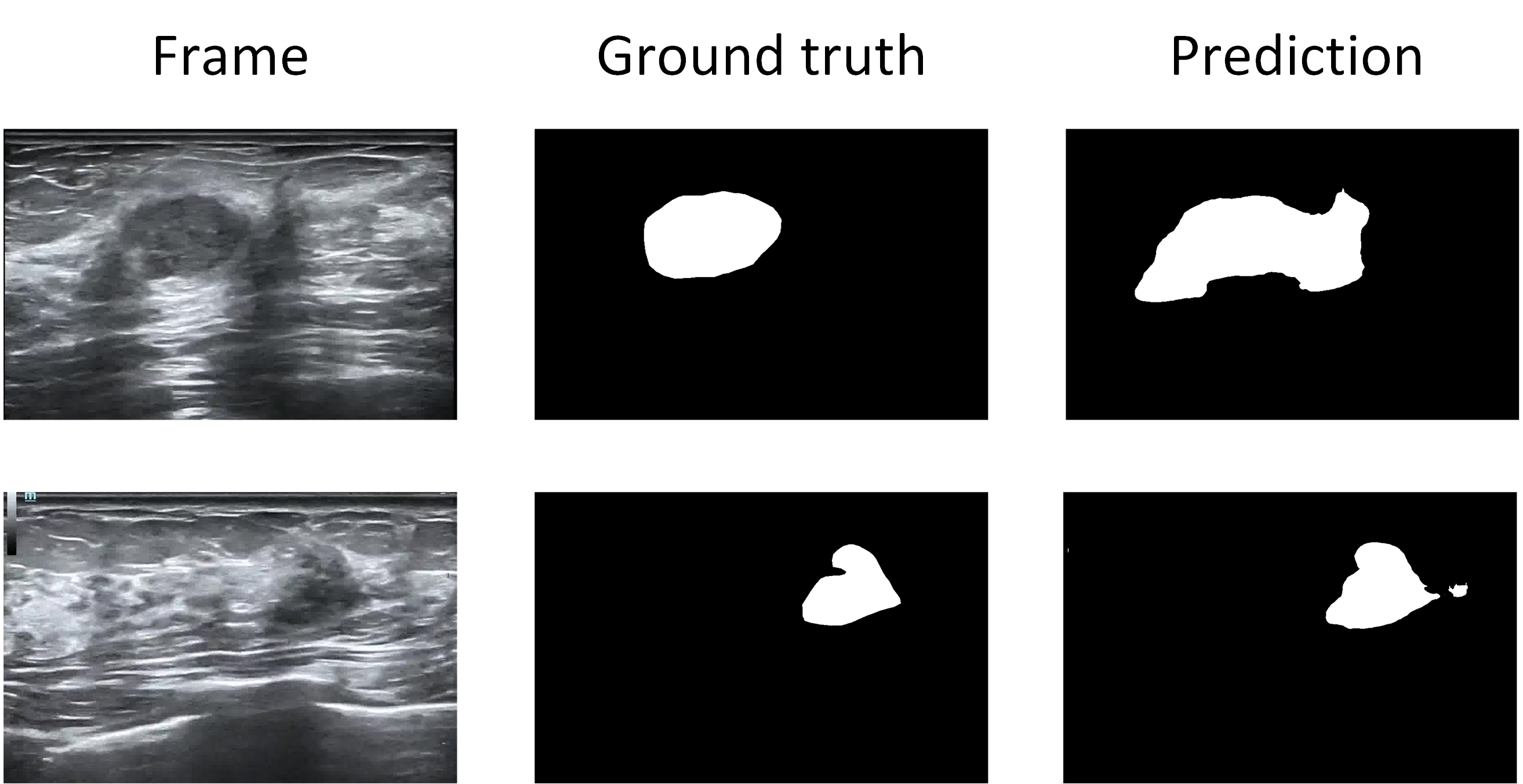}
    \caption{Challenging samples. The table consists of three columns: the first column displays frames used for prediction, the second column shows ground truth, and the third column represents the prediction results.}
    \label{fig6}
\end{figure}

\section{DISCUSSION}

We introduce a novel Spatial-Temporal Progressive Fusion Network, which effectively addresses the problem of incorrect segmentation of non-lesion regions and accurately segments blurred boundaries.
However, some misjudgements still occur in some cases.
For example, when there are shadows around a lesion, the network may incorrectly segment some of the shadows as lesion areas, resulting in inaccurate boundary, as shown in the first row of Fig.~\ref{fig6}.
In addition, introducing prior knowledge from the previous frame successfully mitigates misjudgements for non-lesion regions, still producing incorrect segmentation when the lesion is very similar to the surrounding background and close to the lesion, as shown in the second row of Fig.~\ref{fig6}. 
Thus, in our future studies, we will go on deeper insights into the pathological characteristics of breast lesions and try to achieve more precise segmentation.
We will explore new approaches in feature extraction and new architectures in enhancing differentiation between the lesion and the surrounding background regions.

In comparison to the work~\cite{b17}, we adopt STM as the baseline and consider utilising the information from the previous frame.
However, in this study~\cite{b17}, the main point is on the fusion of the previous frame and the mask of the previous frame, thus fully exploiting the role of the previous frame in space.
In contrast, we map the mask of the previous frame onto the current frame as a prior knowledge for suppressing cluttered information in the blurred background, especially those regions that are similar to the lesion.
In addition, we consider further fusion of more detailed information in the process of refining the boundaries to obtain more clear detailed information and to make the boundaries more accurately.

\section{CONCLUSION}

In this paper, we propose a Spatial-Temporal Progressive Fusion network (STPFNet) to solve the problems of blurred boundary and irregular shape of ultrasound breast lesions.
We first make full use of spatial-temporal information through the temporal fusion module and the spatial fusion module. Then, we perform multi-scale fusion to fuse temporal and spatial features as well as features from the encoder to get more detailed information. We also use the previous frame as a prior knowledge to locate the lesion area.
Finally, we construct a new UVBLS200 dataset for breast lesion segmentation. 
We perform a comparative evaluation of our proposed method against several other state-of-the-art techniques on the UVBLS200 dataset. The results not only demonstrate the effectiveness of our approach compared to other methods, but also indirectly show our challenging dataset.
In the future, we intend to dig deeper into the UVBLS200 dataset to address more challenges in our dataset. 

\section*{Acknowledgment}
Thanks to the First Affiliated Hospital of Anhui Medical University for collecting and annotating the UVBLS200 dataset.


\begin{thebibliography}{00}
\bibitem{b1}
Y. Shu, H. Li, B. Xiao, X. Bi, and W. Li, “Cross-Mix Monitoring for Medical Image Segmentation With Limited Supervision,” \textit{IEEE Transactions on Multimedia}, pp. 1700–1712, Jan. 2023, doi: 10.1109/tmm.2022.3154159.
\bibitem{b2} 
AhmedM. Alaa, KyeongH. Moon, W. Hsu, and M. Schaar, “ConfidentCare: A Clinical Decision Support System for Personalized Breast Cancer Screening,” \textit{IEEE Transactions on Multimedia,IEEE Transactions on Multimedia}, Feb. 2016.
\bibitem{b3} 
O. M. Bucci, L. Crocco, and R. Scapaticci, “On the Optimal Measurement Configuration for Magnetic Nanoparticles-Enhanced Breast Cancer Microwave Imaging,” \textit{IEEE Transactions on Biomedical Engineering}, pp. 407–414, Feb. 2015, DOI: 10.1109/tbme.2014.2355411.
\bibitem{b4} 
H. Hille, “Advances in Breast Ultrasound,” in \textit{Sonography}, 2012. DOI: 10.5772/30078.
\bibitem{b5} 
A. Jalalian, S. B. T. Mashohor, H. R. Mahmud, M. I. B. Saripan, A. R. B. Ramli, and B. Karasfi, “Computer-aided detection/diagnosis of breast cancer in mammography and ultrasound: a review,” \textit{Clinical Imaging}, pp. 420–426, May 2013, DOI: 10.1016/j.clinimag.2012.09.024.
\bibitem{b6}
Y. Xu, Y. Wang, J. Yuan, Q. Cheng, X. Wang, and P. L. Carson, “Medical breast ultrasound image segmentation by machine learning,” \textit{Ultrasonics}, pp. 1–9, Jan. 2019, DOI: 10.1016/j.ultras.2018.07.006.
\bibitem{b7} 
Y. Shu, H. Li, B. Xiao, X. Bi, and W. Li, “Cross-Mix Monitoring for Medical Image Segmentation With Limited Supervision,” \textit{IEEE Transactions on Multimedia}, pp. 1700–1712, Jan. 2023, doi: 10.1109/tmm.2022.3154159.
\bibitem{b8} 
S. Pradeep and P. Nirmaladevi, “A Review on Speckle Noise Reduction Techniques in Ultrasound Medical images based on Spatial Domain, Transform Domain and CNN Methods,” \textit{IOP Conference Series: Materials Science and Engineering}, p. 012116, Feb. 2021, DOI: 10.1088/1757-899x/1055/1/012116.
\bibitem{b9} 
J. Liu \textit{et al.}, “Speckle noise reduction for medical ultrasound images based on cycle-consistent generative adversarial network,”\textit{ Biomedical Signal Processing and Control}, vol. 86, p. 105150, Sep. 2023, DOI: 10.1016/j.bspc.2023.105150.
\bibitem{b10}
H. Wu, X. Huang, X. Guo, Z. Wen and J. Qin, “Cross-Image Dependency Modeling for Breast Ultrasound Segmentation,” in IEEE Transactions on Medical Imaging, vol. 42, no. 6, pp. 1619-1631, June 2023, doi: 10.1109/TMI.2022.3233648.
\bibitem{b11}
G.-Q. Zhou \emph{et al.}, “A Single-Shot Region-Adaptive Network for Myotendinous Junction Segmentation in Muscular Ultrasound Images,” \textit{IEEE Transactions on Ultrasonics, Ferroelectrics, and Frequency Control}, vol. 67, no. 12, pp. 2531-2542, Dec. 2020, DOI: 10.1109/tuffc.2020.2979481.
\bibitem{b12}
A. Ouahabi and A. Taleb-Ahmed, “RETRACTED: Deep learning for real-time semantic segmentation: Application in ultrasound imaging,” \textit{Pattern Recognition Letters}, pp. 27-34, Apr. 2021, DOI: 10.1016/j.patrec.2021.01.010.
\bibitem{b13}
Y. Zhou \emph{et al.}, “Multi-task learning for segmentation and classification of tumors in 3D automated breast ultrasound images,” \textit{Medical Image Analysis}, pp. 101918, May 2021, DOI: 10.1016/j.media.2020.101918.
\bibitem{b14}
K. Wang, S. Liang, and Y. Zhang, “Residual Feedback Network for Breast Lesion Segmentation in Ultrasound Image,” in \textit{Medical Image Computing and Computer Assisted Intervention - MICCAI 2021},Lecture Notes in Computer Science, 2021, pp. 471-481.
\bibitem{b15}
C. Xue \emph{et al.}, “Global guidance network for breast lesion segmentation in ultrasound images,” \textit{Medical Image Analysis}, pp. 101989, May 2021, DIO: 10.1016/j.media.2021.101989.
\bibitem{b16}
Y. Kim, “Convolutional Neural Networks for Sentence Classification,” in \textit{Proceedings of the 2014 Conference on Empirical Methods in Natural Language Processing (EMNLP)}, Doha, Qatar, Jan. 2014. DOI: 10.3115/v1/d14-1181.
\bibitem{b17}
Li, J., Zheng, Q., Li, M., Liu, P., Wang, Q., Sun, L., Zhu, L., “Rethinking Breast Lesion Segmentation in Ultrasound: A New Video Dataset and A Baseline Network,” in \textit{Medical Image Computing and Computer Assisted Intervention - MICCAI 2022.} pp. 391-400. 
\bibitem{b18}
P. Jiang, J. Peng, G. Zhang, E. Cheng, V. Megalooikonomou, and H. Ling, “Learning-based automatic breast tumor detection and segmentation in ultrasound images,” in \textit{2012 9th IEEE International Symposium on Biomedical Imaging (ISBI)}, Barcelona, Spain, May 2012. DOI: 10.1109/isbi.2012.6235878.
\bibitem{b19}
L. Cai and Y. Wang, “A phase-based active contour model for segmentation of breast ultrasound images,” in \textit{2013 6th International Conference on Biomedical Engineering and Informatics}, Hangzhou, China, Dec. 2013. DOI: 10.1109/bmei.2013.6746913.
\bibitem{b20}
F. Kharajinezhadian, F. Yazdani, P. P. Isfahani, and M. Kavousi, “Automatic Breast Tumor Classification in Ultrasound Images Using Morphological Features and New Texture Analysis Based on Image Visibility Graph and Gabor Filters,” \textit{SN Computer Science}, Oct. 2022, DOI: 10.1007/s42979-022-01431-3.
\bibitem{b21}
Q. Huang, Y. Luo, and Q. Zhang, “Breast ultrasound image segmentation: a survey,” \textit{International Journal of Computer Assisted Radiology and Surgery}, pp. 493-507, Mar. 2017, DOI: 10.1007/s11548-016-1513-1.
\bibitem{b22}
S. U. Khan, N. Islam, Z. Jan, K. Haseeb, S. I. A. Shah, and M. Hanif, “A machine learning-based approach for the segmentation and classification of malignant cells in breast cytology images using gray level co-occurrence matrix (GLCM) and support vector machine (SVM),” \textit{Neural Computing and Applications}, pp. 8365-8372, Jun. 2022, DOI: 10.1007/s00521-021-05697-1.
\bibitem{b23}
H. Dhahri, E. Al Maghayreh, A. Mahmood, W. Elkilani, and M. Faisal Nagi, “Automated Breast Cancer Diagnosis Based on Machine Learning Algorithms,” \textit{Journal of Healthcare Engineering}, vol. 2019, pp. 1-11, Nov. 2019, DOI: 10.1155/2019/4253641.
\bibitem{b24}
R. Almajalid, J. Shan, Y. Du, and M. Zhang, “Development of a Deep-Learning-Based Method for Breast Ultrasound Image Segmentation,” in 2018 17th \textit{IEEE International Conference on Machine Learning and Applications (ICMLA)}, Orlando, FL, Dec. 2018. DOI: 10.1109/icmla.2018.00179.
\bibitem{b25}
Y. Hu \emph{et al.}, “Automatic tumor segmentation in breast ultrasound images using a dilated fully convolutional network combined with an active contour model,” \textit{Medical Physics}, pp. 215-228, Jan. 2019, DOI: 10.1002/mp.13268.
\bibitem{b26}
Z. Ning, S. Zhong, Q. Feng, W. Chen, and Y. Zhang, “SMU-Net: Saliency-Guided Morphology-Aware U-Net for Breast Lesion Segmentation in Ultrasound Image,” \textit{IEEE Transactions on Medical Imaging}, pp. 476–490, Feb. 2022, doi: 10.1109/tmi.2021.3116087.
\bibitem{b27}
S. W. Cho, N. R. Baek, and K. R. Park, “Deep Learning-based Multi-stage Segmentation Method Using Ultrasound Images for Breast Cancer Diagnosis,” \textit{Journal of King Saud University - Computer and Information Sciences}, vol. 34, no. 10, pp. 10273–10292, Nov. 2022, DOI: 10.1016/j.jksuci.2022.10.020.1.
\bibitem{b28}
O. Ronneberger, P. Fischer, and T. Brox, “U-Net: Convolutional Networks for Biomedical Image Segmentation,” in \textit{Lecture Notes in Computer Science, Medical Image Computing and Computer-Assisted Intervention - MICCAI 2015}, 2015, pp. 234-241. 
\bibitem{b29}
H. Soltani, M. Amroune, I. Bendib, and M. Y. Haouam, “Breast Cancer Lesion Detection and Segmentation Based On Mask R-CNN,” in \textit{2021 International Conference on Recent Advances in Mathematics and Informatics (ICRAMI)}, Tebessa, Algeria, Sep. 2021. DOI: 10.1109/icrami52622.2021.9585913.
\bibitem{b30}
Wei Wei, K. N. Ngan, and A. Habili, “Multiple feature clustering algorithm for automatic video object segmentation,” in \textit{2004 IEEE International Conference on Acoustics}, Speech, and Signal Processing, Montreal, Que., Canada, Sep. 2004. DOI: 10.1109/icassp.2004.1326622.
\bibitem{b31}
S. W. Oh, J.-Y. Lee, N. Xu, and S. J. Kim, “Video Object Segmentation Using Space-Time Memory Networks,” in \textit{2019 IEEE/CVF International Conference on Computer Vision (ICCV)}, Seoul, Korea (South), Oct. 2019. DOI: 10.1109/iccv.2019.00932.
\bibitem{b32}
H. Seong, J. Hyun, and E. Kim, “Kernelized Memory Network for Video Object Segmentation,” in \textit{Computer Vision - ECCV 2020},Lecture Notes in Computer Science, 2020, pp. 629–645. DOI: 10.1007/978-3-030-58542-6\_38.
\bibitem{b33}
H. Xie, H. Yao, S. Zhou, S. Zhang, and W. Sun, “Efficient Regional Memory Network for Video Object Segmentation,” in \textit{2021 IEEE/CVF Conference on Computer Vision and Pattern Recognition (CVPR)}, Nashville, TN, USA, Jun. 2021. DOI: 10.1109/cvpr46437.2021.00134.
\bibitem{b34}
K. He, X. Zhang, S. Ren, and J. Sun, “Deep Residual Learning for Image Recognition,” in \textit{2016 IEEE Conference on Computer Vision and Pattern Recognition (CVPR)}, Las Vegas, NV, USA, Jun. 2016. DOI: 10.1109/cvpr.2016.90.
\bibitem{b35}
S. Zhao, Y. Wu, S. Wang, W. Ke, and H. Sheng, “Mask Guided Spatial-Temporal Fusion Network for Multiple Object Tracking,” in \textit{2022 IEEE International Conference on Image Processing (ICIP)}, Bordeaux, France, Oct. 2022. DOI: 10.1109/icip46576.2022.9898054. 
\bibitem{b36}
Z. Zhou, M. M. Rahman Siddiquee, N. Tajbakhsh, and J. Liang, “Unet++: A nested u-net architecture for medical image segmentation,” in \textit{Deep Learning in Medical Image Analysis and Multimodal Learning for Clinical Decision Support},Lecture Notes in Computer Science, 2018, pp. 3-11. DOI: 10.1007/978-3-030-00889-5\_1.
\bibitem{b37}
P. Chao, C.-Y. Kao, Y. Ruan, C.-H. Huang, and Y.-L. Lin, “HarDNet: A Low Memory Traffic Network,” in \textit{2019 IEEE/CVF International Conference on Computer Vision (ICCV)}, Seoul, Korea (South), Oct. 2019. DOI: 10.1109/iccv.2019.00365.
\bibitem{b38}
X. Zhao, L. Zhang, and H. Lu, “Automatic Polyp Segmentation via Multi-scale Subtraction Network,” in\textit{ Medical Image Computing and Computer Assisted Intervention - MICCAI 2021},Lecture Notes in Computer Science, 2021, pp. 120-130. DOI: 10.1007/978-3-030-87193-2\_12.
\bibitem{b39}
N. K. Tomar, A. Shergill, B. Rieders, U. Bagci, and D. Jha, “Transresunet: Transformer based resu-net for real-time colonoscopy polyp segmentation,” \textit{arXiv preprint arXiv:}2206.08985, 2022.
\bibitem{b40}
H. Wang, P. Cao, J. Wang, and O. R. Zaiane, “UCTransNet: Rethinking the Skip Connections in U-Net from a Channel-wise Perspective with Transformer,” \textit{Proceedings of the AAAI Conference on Artificial Intelligence}, pp. 2441-2449, Jul. 2022, DOI: 10.1609/aaai.v36i3.20144.
\bibitem{b41}
Y. Liang, X. Li, NavidH. Jafari, and J. Chen, “Video Object Segmentation with Adaptive Feature Bank and Uncertain-Region Refinement,” \textit{Neural Information Processing Systems,Neural Information Processing Systems}, Jan. 2020.
\bibitem{b42}
M. Zhang et al., “Dynamic Context-Sensitive Filtering Network for Video Salient Object Detection,” in \textit{2021 IEEE/CVF International Conference on Computer Vision (ICCV)}, Montreal, QC, Canada, Oct. 2021. DOI: 10.1109/iccv48922.2021.00158.
\bibitem{b43}
Y. Su, J. Deng, R. Sun, G. Lin, H. Su and Q. Wu, “A Unified Transformer Framework for Group-based Segmentation: Co-Segmentation, Co-Saliency Detection and Video Salient Object Detection,” in \textit{IEEE Transactions on Multimedia}, vol. 26, pp. 313-325, 2024, doi: 10.1109/TMM.2023.3264883.
\bibitem{b44}
Lin, Junhao, et al. "Shifting More Attention to Breast Lesion Segmentation in Ultrasound Videos," in \textit{Medical Image Computing and Computer Assisted Intervention - MICCAI 2023.} pp. 497-507. 

 


\end{thebibliography}
\end{document}